\documentclass[aps,12pt,final,notitlepage,oneside,onecolumn,nobibnotes,nofootinbib,
superscriptaddress,noshowpacs,centertags,amsmath,amssymb]{revtex4}

\usepackage{graphicx}
\usepackage{dcolumn}
\usepackage{bm}

\newcommand{\hs}{\hspace*{0.5cm}}
\newcommand{\vs}{\vspace*{0.5cm}}
\newcommand{\be}{\begin{equation}}
\newcommand{\ee}{\end{equation}}
\newcommand{\bea}{\begin{eqnarray}}
\newcommand{\eea}{\end{eqnarray}}
\newcommand{\bary}{\begin{array}}
\newcommand{\eary}{\end{array}}
\newcommand{\bit}{\begin{itemize}}
\newcommand{\eit}{\end{itemize}}
\newcommand{\ben}{\begin{enumerate}}
\newcommand{\een}{\end{enumerate}}
\newcommand{\crn}{\nonumber \\}

\newcommand{\al}{\alpha}
\newcommand{\la}{\lambda}
\newcommand{\bet}{\beta}
\newcommand{\ga}{\gamma}
\newcommand{\va}{\varphi}

\newcommand{\pa}{\partial}
\newcommand{\fr}{\frac}

\newcommand{\bc}{\begin{center}}
\newcommand{\ec}{\end{center}}

\newcommand{\ep}{\epsilon}

\begin{document}

\title{
 Inflationary scenario in the supersymmetric\\
economical 3-3-1 model}

\author{ Do T. Huong} %
\email{dthuong@iop.vast.ac.vn}
\author{ Hoang N. Long }%
 \email{hnlong@iop.vast.ac.vn}
\affiliation{%
Institute of Physics,
Hanoi, Vietnam
}%

\date{\today}

\begin{abstract}
We construct  the supersymmetric economical 3-3-1 model which
contains inflationary scenario and avoids the  monopole puzzle.
Based on the spontaneous symmetry breaking pattern (with three
steps), the $F$-term inflation is derived. The slow-roll
parameters $\ep$ and $\eta$ are calculated. By imposing as
experimental five-year WMAP data on the spectral index $n$, we
have derived a  constraint on the number of e-folding $N_Q$ to be
in the range from 25 to 50. The scenario for large-scale structure
formation implied by the model is a mixed scenario for inflation
and cosmic string, and the contribution to the CMBR temperature
anisotropy depends  on the ratio $\frac{M_X}{M_P}$. From the COBE
data, we have obtained the constraint on the $M_X$ to be
$M_X\in[1.22 \times 10^{16} - 0.98 \times 10^{17}]$ GeV. The upper
value $M_X\simeq 10^{17}$ GeV is a result of the analysis in which
the inflationary contribution to the temperature fluctuations
measured
 by the COBE is 90\%.  The coupling $\alpha$ varies in the
range: $ 10^{-7} - 10^{-1} $. This value is not so small, and  it
is a common characteristics  of the supersymmetric unified models
with  the inflationary scenario. The spectral index $n$ is a
little bit smaller than $0.98$. The SUGRA corrections  are
slightly different from the previous consideration. When $\xi \ll
1$ and $ \al $ lies in the above range, the spectral index gets
the value consistent with the experimental five-year WMAP data.
Comparing with string theory, one gets $\xi < 10^{-8}$. Numerical
analysis shows that $ \al \approx  10^{-6}$. To get inflation
contribution to the CMBR temperature anisotropy $\approx 90 \%$,
the mass scale $M_X \sim 3.5 \times 10^{14}$ GeV.

\end{abstract}


\maketitle

\section{\label{intro}Introduction}

 This time is a golden age of cosmology and astrophysics. Many
abstractive notions such as Black Holes, Dark matter, etc. step by
step become more popular and widely  accepted subjects. In the
past, cosmology often relied on philosophical or aesthetic
arguments; now it is maturing to become an exact science. The
1990s have seen consolidation of theoretical cosmology, coupled
with dramatical observational advances, including the emergence of
an entirely new field of observational astronomy - the study of
irregularities in the microwave background radiation. A key idea
of modern cosmology is cosmological
inflation~\cite{infsce,lind383}, which is a possible theory of the
origin of all structures in the Universe, including ourselves!

By the way, the rapid development of elementary particle theory
has not only led to great advances in our understanding of
particle interactions at superhigh energies, but also to
significant progress in the theory of superdense matter. The
Standard Model (SM) of strong and electroweak interactions was
obtained within the scope of gauge theories with spontaneous
symmetry breaking. For the first time, it became possible to
investigate strong and weak interaction processes using high-order
perturbation theory. A remarkable property of these
theories-asymptotic freedom-also made it possible in principle to
describe interactions of elementary particles up to center-of-mass
energies $E \sim M_P \sim 10^{19}$ GeV, that is, up to the Planck
energy, where quantum gravity effects become important.

This result comprised the first evidence for the importance of
unified theories of  elementary particles and the theory of
superdense matter for the development of the theory of the
evolution of the universe. Up to mid-1960's, it was still not
clear whether the early universe had been hot or cold. The
critical juncture marking the beginning of the second stage in the
development of modern cosmology was discovery of the 2.7$K$
microwave background radiation arriving from the farthest reaches
of the universe. The existence of the microwave background had
been predicted by the hot-universe theory, which gained immediate
and widespread acceptance after the discovery.

However, there are a lot of difficulties (see, for
example,~\cite{linde1}) in modern cosmology such as flatness,
horizon, primordial monopole problems,  etc. It is all the more
surprising, then, that many of these problems, together with a
number of others that predate the hot universe theory, have been
resolved in the context of one fairly simple scenario for the
development of the universe - the so-called inflationary universe
scenario~\cite{infsce}. Inflation assumes that there was a period
in the very early universe when the potential  and vacuum energy
density
dominated the energy of the universe, so that the cosmic scale
factor grew exponentially.

The important ingredient of the inflationary  scenario is a scalar
field $\va$ having effective potential $V(\va)$ with some
properties (satisfying  many constrains that are rather
unnatural). This scalar field is called inflaton.  It was found
that  many unified theories contain the mentioned inflaton. Up to
date, the inflationary scenario has been considered in  the
framework of the models such as:  supersymmetric $SU(5)$,
extra-dimensional, superstring, etc. Recently a general scenario
for unification of dark matter and inflation into a single field
has been proposed~\cite{darkinf}.

Nevertheless, in building supersymmetric grand unified models
intended to be consistent with cosmology one is confronted with
two main problems. The first problem is that any semisimple grand
unified gauge group, which is broken down to the SM
 $\mathrm{SU}(3)_C \otimes \mathrm{SU}(2)_L \otimes
 \mathrm{U}(1)_Y$, inevitably leads to the formation of topologically stable
monopoles, according to the Kibble mechanism~\cite{kibb}. These
monopole, if present today, would dominate the energy density of
the universe, and our universe would be different from what we
observe. Even if the grand unified gauge group $G$ is not
semisimple, it may still be confronted with the monopole problem.
The second problem, which is directly related to this first one,
is that inflation usually requires
  tough fine-tuning of the parameters. This problem and the
previously mentioned one are resolved in the inflation scenario.

One of the greatest triumphs of physics in the  twentieth  century
is the Standard Model, which  provides a remarkable successful
description of presently known phenomena. In spite of these
successes, it fails to explain several fundamental issues like
generation number puzzle, neutrino masses and oscillations, the
origin of charge quantization, CP violation, etc.  One of the
simplest solutions to these problems is to enhance the SM symmetry
$\mathrm{SU}(3)_{C} \otimes \mathrm{SU}(2)_{L} \otimes
\mathrm{U}(1)_{Y}$ to $\mathrm{SU}(3)_{C} \otimes
\mathrm{SU}(3)_{L} \otimes \mathrm{U}(1)_{X}$ (called 3-3-1 for
short)~\cite{ppf,flt,331rh} gauge group.
 One of the main motivations to study this kind of models
is an explanation in part of the generation number ($n_f$)puzzle.
In the 3-3-1 models, each generation is not anomaly free; and the
model becomes anomaly free if one of quark families behaves
differently from other two. Consequently, the number of
generations is multiple of the color number. Combining with the
QCD asymptotic freedom   which requires $n_f < 5$, the  number of
generations  has to be three.

In one of the 3-3-1 models, the right-handed neutrinos are in
bottom of the lepton triplets \cite{331rh} and three Higgs
triplets are required. It is worth noting that there are two Higgs
triplets with {\it neutral components in the top and bottom}. In
the earlier version, these triplets can have vacuum expectation
value (VEV) either on the top or in the bottom, but not in both.
Assuming that all neutral components in the triplet can have VEVs,
we are able to reduce number of triplets in the model to be two
~\cite{ponce,haihiggs} (for a review, see ~\cite{ahep}). Such a
scalar sector is minimal, therefore it has been called the
economical 3-3-1 model~\cite{higgseconom}. In a series of papers,
we have developed and proved that this non-supersymmetric version
is consistent, realistic, and very rich in physics
\cite{haihiggs,higgseconom, dlhh,dls1}.

On the other hand, a triviality of the unification among  the
internal $G$ and external $P$ space-time symmetries can be avoided
by new symmetry called supersymmetry~\cite{susy,martin}.
 One of the intriguing
features of supersymmetric theories is that the Higgs spectrum
(unfortunately, the only part of the SM
 still not discovered)
is quite constrained~\cite{shiggs}.

A supersymmetric version of the minimal version (without extra
lepton) has been constructed in ~\cite{msusy} and its scalar
sector was  studied in ~\cite{duongma}. Lepton masses in framework
of the above-mentioned model were presented in
\cite{leptonmassm331}, while potential discovery of supersymmetric
particles  was studied in \cite{consm331}. In \cite{longpal}, the
$R$-parity violating interaction was applied for  instability of
the proton.

The supersymmetric version of the 3-3-1 model with right-handed
neutrinos has already been constructed in~\cite{s331r}. The scalar
sector was considered in~\cite{scalarrhn} and neutrino mass was
studied in ~\cite{marcos}.

 A supersymmetric version of the economical 3-3-1 model has been
constructed in  \cite{susyeco}. Some interesting features such as
Higgs bosons with masses equal to that of the gauge bosons: the
$W$ ($m^2_{\varrho_{1}^+}=m^2_W$) and the bileptons $X$ and $Y$
($m^2_{\zeta^\pm_4} = m^2_Y$), have been pointed out in
\cite{higph}. Sfermions in this model have been considered in
\cite{jhep}. In \cite{jhep2} we have shown that bino-like
neutralino can be candidate for dark matter (DM).

The aim of the present paper is to show that the recently
constructed supersymmetric economical 3-3-1 model contains the
necessary inflation. It is emphasized that to have the
inflationary scenario, we have to do  spontaneous  symmetry
breakdown by an unusual  way, namely through three steps instead
of two as in the previous works~\cite{susyeco,higph}.  In
\cite{Jen}, the $D$-term inflation has been considered. The
alternative $F$-term inflation is a subject of  the present paper.

This article is organized as follows. In Section ~\ref{parcontent}
we present fermion and scalar content in the supersymmetric
economical 3-3-1 model. The necessary parts of Lagrangian are
given in Section~\ref{Lagrang}. Section~\ref{effective} is devoted
for the effective potential in the model with inflation.
 In Section~\ref{finf} the $F$-term inflation is considered
 and  slow roll parameters such as $\ep, \eta$ and
 spectral index $n$ are
calculated and constrained by the WMAP data.  Section \ref{sec6}
is devoted for the standard $F$ term inflation with minimal
K\"{a}hler potential.  We summary our results and make conclusions
in the  section~\ref{conclusion}.

\section{\label{parcontent}Particle content
 and spontaneous  symmetry breakdown}

To proceed further, the necessary features of the supersymmetric
economical 3-3-1 model \cite{haihiggs} will be presented. The
superfield content in the present paper is defined in a standard
way as follows \be \widehat{F}= (\widetilde{F}, F),\hs \widehat{S}
= (S, \widetilde{S}),\hs \widehat{V}= (\lambda,V), \ee where the
components $F$, $S$ and $V$ stand for the fermion, scalar, and
vector fields of the economical 3-3-1 model, while their
superpartners are denoted as $\widetilde{F}$, $\widetilde{S}$ and
$\lambda$, respectively \cite{susy,s331r}.

The superfields for the leptons under the 3-3-1 gauge group
transform as
\begin{equation}
\widehat{L}_{a L}=\left(\widehat{\nu}_{a}, \widehat{l}_{a},
\widehat{\nu}^c_{a}\right)^T_{L} \sim ( {\bf 1},{\bf 3},-1/3),\hs
  \widehat {l}^{c}_{a L} \sim ({\bf 1},{\bf 1},1),\label{l2}
\end{equation} where $\widehat{\nu}^c_L=(\widehat{\nu}_R)^c$ and $a=1,2,3$
is a generation index.

It is worth mentioning that in the economical version the first
generation of quarks should be different from others \cite{dlhh}.
The superfields for the left-handed quarks of the first generation
are in triplets: \be \widehat Q_{1L}= \left(\widehat { u}_1,\
                        \widehat {d}_1,\
                        \widehat {u}^\prime
 \right)^T_L \sim ( {\bf 3},{\bf 3},1/3),\label{quarks3}\ee
where the right-handed singlet counterparts are given by\be
\widehat {u}^{c}_{1L},\ \widehat { u}^{ \prime c}_{L} \sim ( {\bf
3^{*}},{\bf 1},-2/3),\hs \widehat {d}^{c}_{1L} \sim ( {\bf
3^{*}},{\bf 1},1/3 ). \label{l5} \ee Conversely, the superfields
for the last two generations transform as antitriplets:
\begin{equation}
\begin{array}{ccc}
 \widehat{Q}_{\alpha L} = \left(\widehat{d}_{\alpha}, - \widehat{u}_{\alpha},
 \widehat{d^\prime}_{\alpha}\right)^T_{L} \sim (
 {\bf 3},{\bf 3^*},0), \hs \al=2,3, \label{l3}
\end{array}
\end{equation}
where the right-handed counterparts are in singlets:
\begin{equation}
\widehat{u}^{c}_{\alpha L} \sim \left( {\bf 3^{*}},{\bf 1},-2/3
\right),\hs \widehat{d}^{c}_{\alpha L},\ \widehat{d}^{\prime
c}_{\alpha L} \sim \left( {\bf 3^{*}},{\bf 1},1/3 \right).
\label{l4}
\end{equation}
The prime superscripts on usual quark types ($u'$ with the
electric charge $q_{u'}=2/3$ and $d'$ with $q_{d'}=-1/3$) indicate
that those quarks are  exotic ones. The mentioned fermion content,
which belongs to that of the 3-3-1 model with right-handed
neutrinos \cite{331rh,haihiggs}, is, of course,  free from
anomaly.

The two  (a minimal number) superfields $\widehat{\chi}$ and
$\widehat {\rho} $ are introduced to span the scalar sector of the
economical 3-3-1 model \cite{higgseconom}: \bea \widehat{\chi}&=&
\left ( \widehat{\chi}^0_1, \widehat{\chi}^-, \widehat{\chi}^0_2
\right)^T\sim ({\bf 1}, {\bf 3},-1/3), \label{l7}\\
\widehat{\rho}&=& \left (\widehat{\rho}^+_1, \widehat{\rho}^0,
\widehat{\rho}^+_2\right)^T \sim  ({\bf 1}, {\bf 3},2/3).
\label{l8} \eea To cancel the chiral anomalies of higgsino sector,
the two extra superfields $\widehat{\chi}^\prime$ and $\widehat
{\rho}^\prime $ must be added as follows: \bea
\widehat{\chi}^\prime&=& \left (\widehat{\chi}^{\prime 0}_1,
\widehat{\chi}^{\prime +},\widehat{\chi}^{\prime 0}_2
\right)^T\sim ({\bf 1}, {\bf 3}^*,1/3),
\label{l9}\\
\widehat{\rho}^\prime &=& \left (\widehat{\rho}^{\prime -}_1,
  \widehat{\rho}^{\prime 0},  \widehat{\rho}^{\prime -}_2
\right)^T\sim ({\bf 1}, {\bf 3}^*,-2/3). \label{l10} \eea

In this model, the $ \mathrm{SU}(3)_L \otimes \mathrm{U}(1)_X$
gauge group is broken via two steps:
 \be \mathrm{SU}(3)_L \otimes
\mathrm{U}(1)_X \stackrel{w,w'}{\longrightarrow}\ \mathrm{SU}(2)_L
\otimes \mathrm{U}(1)_Y\stackrel{v,v',u,u'}{\longrightarrow}
\mathrm{U}(1)_{Q},\label{stages}\ee where the VEVs are defined by
\bea
 \sqrt{2} \langle\chi\rangle^T &=& \left(u, 0, w\right), \hs \sqrt{2}
 \langle\chi^\prime\rangle^T = \left(u^\prime,  0,
 w^\prime\right),\\
\sqrt{2}  \langle\rho\rangle^T &=& \left( 0, v, 0 \right), \hs
\sqrt{2} \langle\rho^\prime\rangle^T = \left( 0, v^\prime,  0
\right).\eea

The vector superfields $\widehat{V}_c$, $\widehat{V}$ and
$\widehat{V}^\prime$ containing the usual gauge bosons are,
respectively, associated with the $\mathrm{SU}(3)_C$,
$\mathrm{SU}(3)_L$, and $\mathrm{U}(1)_X $ group factors. The
color and flavor vector superfields have expansions in the
Gell-Mann matrix bases $T^a=\lambda^a/2$ $(a=1,2,...,8)$ as
follows:\bea \widehat{V}_c &=& \fr{1}{2}\lambda^a
\widehat{V}_{ca},\hs
\widehat{\overline{V}}_c=-\fr{1}{2}\lambda^{a*}
\widehat{V}_{ca};\hs \widehat{V} = \fr{1}{2}\lambda^a
\widehat{V}_{a},\hs \widehat{\overline{V}}=-\fr{1}{2}\lambda^{a*}
\widehat{V}_{a},\eea where an overbar $^-$ indicates complex
conjugation. For the vector superfield associated with
$\mathrm{U}(1)_X$, we normalize as follows: \be X \hat{V}'= (XT^9)
\hat{B}, \hs T^9\equiv\fr{1}{\sqrt{6}}\mathrm{diag}(1,1,1).\ee In
the following, we are denoting the gluons by $g^a$ and their
respective gluino partners by $\lambda^a_{c}$, with $a=1,
\ldots,8$. In the electroweak sector, $V^a$ and $B$ stand for the
$\mathrm{SU}(3)_{L}$ and $\mathrm{U}(1)_{X}$ gauge bosons with
their gaugino partners $\lambda^a_{V}$ and $\lambda_{B}$,
respectively.

\section {\label{Lagrang}The models}
With the superfields  given above, we can now construct the
supersymmetric economical 3-3-1 model containing the Lagrangians:
$\mathcal{L}_{susy}+\mathcal{L}_{soft}$, where the first term is
supersymmetric part, whereas the last term  explicitly breaks  the
supersymmetry.

\subsection{\label{suplagran}  Supersymmetric model  without
inflationary scenario}

The supersymmetric Lagrangian can be decomposed into four relevant
parts:
\begin{eqnarray} \mathcal{L}_{susy} &=& {\mathcal L}_{gauge}
+{\mathcal L}_{lepton} +{\mathcal L}_{quark} +{\mathcal
L}_{scalar}. \label{lg2}
\end{eqnarray}
The first term has the form
\begin{eqnarray}
{\mathcal L}_{gauge} &=& \frac{1}{4} \int
d^{2}\theta\;\mathcal{W}_{ca} \mathcal{W}_{ca}+ \frac{1}{4} \int
d^{2}\theta\;\mathcal {W}_a \mathcal{W}_a+ \frac{1}{4} \int
d^{2}\theta \mathcal{W}^{\prime}\mathcal{W}^{ \prime}+  \crn && +
\frac{1}{4} \int
d^{2}\bar{\theta}\;\overline{\mathcal{W}}_{ca}\overline{\mathcal{W}}_{ca}+
\frac{1}{4} \int d^{2}\bar{\theta}\;\overline{\mathcal{W}}_a
\overline{\mathcal{W}}_a+ \frac{1}{4} \int  d^{2}\bar{\theta}
\overline{\mathcal{W}}^{ \prime}\overline{\mathcal{W}}^{
\prime},\label{gaug}
\end{eqnarray}
where the chiral superfields $\mathcal{W}_{c}$, $\mathcal{W}$, and
$\mathcal{W}^{ \prime}$ are defined by
\begin{eqnarray}
\mathcal{W}_{c\zeta}&=&- \frac{1}{8g_s} \bar{D} \bar{D} e^{-2g_s
\hat{V}_{c}}
D_{\zeta} e^{2g_s \hat{V}_{c}},\nonumber \\
\mathcal{W}_{\zeta}&=&- \frac{1}{8g} \bar{D} \bar{D} e^{-2g
\hat{V}}
D_{\zeta} e^{2g \hat{V}}, \nonumber \\
\mathcal{W}^{\prime}_{\zeta}&=&- \frac{1}{4} \bar{D} \bar{D}
D_{\zeta} \hat{V}^{\prime}, \,\ \zeta=1,2, \label{cforca}
\end{eqnarray}
with the gauge couplings $g_{s}$, $g$, and $g^{\prime}$ respective
to $\mathrm{SU}(3)_C$, $\mathrm{SU}(3)_L$, and $\mathrm{U}(1)_X$.
The $D_{\zeta}$ and $\bar{D}_{\dot{\zeta}}$ are the chiral
covariant derivatives of SUSY algebra as presented in \cite{susy}.

The second and third terms are given by \begin{eqnarray} {\mathcal
L}_{lepton} &=& \int d^{4}\theta \left[ \hat{\bar{L}}_{a
L}e^{2(g\hat{V}- \frac{g^\prime}{3}\hat{V}^\prime)}\hat{L}_{a L}
+\hat{\bar{l}}^c_{a L}e^{2g^\prime \hat{V}^\prime}\hat{l}^c_{a L}
\right] \label{lg3}
\end{eqnarray}
 and
\begin{eqnarray}
{\mathcal L}_{quark} &=& \int d^{4}\theta \left[ \hat{\bar{Q}}_{1
L}e^{{2(g_s\hat{V}_c+g\hat{V}+\frac{g^\prime}{3}
\hat{V}^\prime)}}\hat{Q}_{1 L} + \hat{\bar{Q}}_{\alpha
L}e^{{2(g_s\hat{V}_c+g\hat{\bar{V}})}}\hat{Q}_{\alpha L}+
\right.\crn  &&+\hat{\bar{u}}^c_{iL}e^{2(g_s
\hat{\bar{V}}_c-\frac{2g^\prime}{3}\hat{V}^\prime) }\hat{u}^c_{iL}
+\hat{\bar{d}}^c_{iL}e^{2(g_s
\hat{\bar{V}}_c+\frac{g^\prime}{3}\hat{V}^\prime)}\hat{d}^c_{iL}+
\nonumber \\ &&+\left. \hat{\bar{u}}^{\prime c}_{L}e^{2(g_s
\hat{\bar{V}}_c-\frac{2g^\prime}{3}\hat{V}^\prime
)}\hat{u}^{\prime c}_{L}+\hat{\bar{d}}^{\prime c}_{\alpha
L}e^{2(g_s
\hat{\bar{V_c}}+\frac{g^\prime}{3}\hat{V^\prime})}\hat{d}^{\prime
c}_{\alpha L} \right]\label{lg3m}
\end{eqnarray}
Finally, the last term can be written as
\begin{eqnarray}
{\mathcal L}_{scalar} &=& \int d^{4}\theta\;\left\{\, \hat{ \bar{
\chi}}e^{2[g\hat{V}+g^{\prime} \left( - \frac{1}{3}\right)
\hat{V}^{\prime}]} \hat{ \chi} + \hat{ \bar{
\rho}}e^{2[g\hat{V}+g^{\prime} \left( \frac{2}{3}\right)
\hat{V}^{\prime}]} \hat{ \rho} + \hat{ \bar{ \chi}}^{\prime}
e^{2[g\hat{ \bar{V}}+g^{\prime} \left( \frac{1}{3}\right)
\hat{V}^{\prime}]}
\hat{ \chi}^{\prime}+\right. \nonumber \\
&&+ \left. \, \hat{ \bar{ \rho}}^{\prime} e^{2[g\hat{
\bar{V}}+g^{\prime} \left( - \frac{2}{3}\right) \hat{V}^{\prime}]}
\hat{ \rho}^{\prime} \right\} +\left( \int d^2 \theta W +
h.c\right) \! \hspace{2mm} \label{esc}
\end{eqnarray}
 with \begin{equation}
W= \frac{W_{2}}{2}+ \frac{W_{3}}{3},
\end{equation}
where $W_2$ and $W_2$ were given in~\cite{susyeco}.

As in~\cite{marcos}, it is useful to impose $R$-parity, which is
determined through the conserved $\mathcal{L}$ and  $\mathcal{B}$
charges (see~\cite{changlong}). Under $R$-parity transformation,
the Higgs and matter superfields change,
respectively~\cite{marcos}:
 \bea
\hat{H}_{1,2}(x,\theta,\bar{\theta}) &\stackrel{{\bf
R}_{d}}{\longmapsto}& \hat{H}_{1,2}(x,-\theta,-\bar{\theta} ),\crn
 \hat{S}(x,\theta,\bar{\theta}) &\stackrel{{\bf
R}_{d}}{\longmapsto}& \hat{S}(x,-\theta,-\bar{\theta} ), \crn
 \hat{\Phi}(x,\theta,\bar{\theta}) &\stackrel{{\bf
R}_{d}}{\longmapsto}& - \hat{\Phi}(x,-\theta,-\bar{\theta} ),
 \ \Phi=Q,u^c,d^c,L,l^c
\label{Rpa1c} \eea Let us separate $W$ into the $R$-parity
conserving ($R$) and violating ($R\!\!\!\!/$) parts~\cite{jhep}:
\be W = W_{R} + W_{R\!\!\!\!/}, \label{Rw} \ee where
 \bea W_{R} & = & \fr 1
2 \left(\mu_{ \chi} \hat{ \chi} \hat{ \chi}^{\prime}+
 \mu_{ \rho} \hat{ \rho} \hat{ \rho}^{\prime}\right)+\crn
 & & + \fr
1 3 \left(\ga_{ab} \hat{L}_{aL} \hat{ \rho}^{\prime}
\hat{l}^{c}_{bL}+ \la^\prime_{ab} \epsilon \hat{L}_{aL}
\hat{L}_{bL}
\hat{\rho}+ \right.\nonumber \\
&&+ \kappa^\prime \hat{Q}_{1L} \hat{\chi}^{\prime} \hat{u}^{\prime
c}_L+ \vartheta_{i}\hat{Q}_{1L} \hat{\rho}^{\prime}
\hat{d}^{c}_{iL} + \pi_{ \alpha i} \hat{Q}_{\alpha
L}\hat{\rho}\hat{u}^{c}_{iL}+ \Pi_{\alpha i} \hat{Q}_{\alpha L}
\hat{\chi}
\hat{d}^{c}_{iL}+ \nonumber \\
&&+ \kappa_{i} \hat{Q}_{1L} \hat{\chi}^{\prime} \hat{u}^{c}_{iL}+
\vartheta^\prime_{ \alpha}\hat{Q}_{1L} \hat{\rho}^{\prime}
\hat{d}^{\prime c}_{\alpha L} +\pi_{\alpha}^{\prime}
\hat{Q}_{\alpha L}\hat{\rho}\hat{u}^{\prime c}_{L} + \left.
\Pi^\prime_{\alpha \beta} \hat{Q}_{\alpha L} \hat{\chi}
\hat{d}^{\prime c}_{\beta L}\right) \label{Rw1} \eea and \bea
W_{R\!\!\!\!/} & = & \fr 1 2 \mu_{0a}\hat{L}_{aL} \hat{
\chi}^{\prime} + \fr 1 3 \left( \la_{a} \epsilon \hat{L}_{aL}
\hat{\chi} \hat{\rho}+ \epsilon
f_{\alpha\beta\gamma}\hat{Q}_{\alpha L} \hat{Q}_{\beta L}
\hat{Q}_{\gamma L}+\right. \crn &&+ \xi_{1i \beta j}
\hat{d}^{c}_{iL} \hat{d}^{\prime c}_{\beta L} \hat{u}^{c}_{j L}+
\xi_{2i \beta } \hat{d}^{c}_{i L} \hat{d}^{\prime c}_{\beta L}
\hat{u}^{\prime c}_{L}+ \xi_{3ijk}
\hat{d}^{c}_{iL} \hat{d}^{c}_{jL} \hat{u}^{c}_{k L} +\nonumber \\
&&+ \xi_{4ij} \hat{d}^{c}_{i L} \hat{d}^{c}_{jL} \hat{u}^{\prime
c}_{L}+ \xi_{5 \alpha \beta i} \hat{d}^{\prime c}_{\alpha L}
\hat{d}^{\prime c}_{\beta L} \hat{u}^{c}_{iL} + \xi_{6 \alpha
\beta} \hat{d}^{\prime c}_{\alpha L}\hat{d}^{\prime c}_{\beta L}
\hat{u}^{\prime c}_{L}+ \crn &&+\left. \xi_{a \alpha
j}\hat{L}_{aL} \hat{Q}_{\alpha L} \hat{d}^{c}_{jL}+
\xi^\prime_{a\alpha \beta}\hat{L}_{aL} \hat{Q}_{\alpha L}
\hat{d}^{\prime c}_{\beta L}\right)\label{wrv}\eea

We remind that as follows from  (\ref{Rpa1c}), the $R\!\!\!\!/$
part contains odd number of {\it matter} superfields.

\subsection{\label{suplagrainf}Supersymmetric model  with the standard hybrid inflation}

   The inflationary mechanism is
currently the most popular model for the origin of structure,
partly because it turns out to give mathematically simple
predictions, but mainly because so far it offers excellent
agreement with the real Universe, such as the microwave
anisotropies. The aim in the present work is to extend the above
supersymmetric version to the one that could be consistent with a
theory of the evolution of the early universe-the model having the
cosmological inflationary scenario. More precisely,  we intend to
construct a hybrid inflationary scheme based on a realistic
supersymmetric
  $ SU(3)_C \otimes SU(3)_L \otimes U(1)_X $
model by adding a singlet superfield $\Phi$ which plays the role
of the inflation, namely the inflaton superfield.

 We remind that the spontaneous symmetry breaking in our model is given
 in (\ref{stages}). This  means that the existence of a $U(1)_Z$ does not
belong to the MSSM  and is spontaneously  broken down  at the
scale $M_X$ by pair of Higgs  superfields   $\chi, \chi^\prime$.
Note that $\chi, \chi^\prime$ are singlets under the MSSM, so they
satisfy the above-mentioned conditions (for details, see
~\cite{Jen}). The inflaton superfield couples with this pair of
Higgs superfields. Therefore, the additional global supersymmetric
renorrmalizable superpotential for the inflation sector is chosen
to be~\cite{infpot,GDvali} \be
W_{inf}(\Phi,\chi,\chi^\prime)=\alpha \Phi \chi \chi^\prime -\mu^2
\Phi.\label{Poten}\ee  The superpotential given by (\ref{Poten})
is the most general potential consistent with a continuous $R$
symmetry under which $\phi \rightarrow e^{i \ga} \phi,  \ W
\rightarrow e^{i \ga} W $, while the product $\chi \chi^\prime$ is
invariant ~\cite{GDvali,linrio}.

 Without loss of generality we can choose $\mu^2,\alpha$ as
 positive real constants by a suitable redefinition of complex
 fields, and the ratio $\frac{\mu} {\sqrt{ \alpha}}$ sets the $U(1)_Z$
 symmetry breaking scale $M_X$. Therefore, the most general
 superpotential consistent with a continuous $R$-symmetry
  is written as
\be W_{tot}=W_R + W_{inf}(\Phi,\chi,\chi^\prime).
\label{Poten1}\ee
 We point out that, at least in the global
supersymmetric case, the $R$-symmetry is the unique choice for
implementing the false vacuum inflationary scenario in a natural
way, i.e, no extra field is needed, apart from the singlet scalar.
It is the only symmetry which can eliminate all of the undesirable
self-coupling of the inflaton $\Phi$, while allowing the linear
term in the superpotential~\cite{GDvali}. With the superpotential
given in (\ref{Poten}), we derive the Higgs scalar potential
 \be V_{tot} =
\Sigma_{i}|F_i|^2 + \frac{1}{2}\sum_{\alpha}|D_\alpha|^2+
V_{soft},\ee where $i$ runs from 1 to the total number of the
chiral superfields in $W_{tot}$, while $V_{soft}$ contains all the
soft terms generated by supersymmetry breaking at the low energy.
The $F$ and $D$ terms are given by
 \be
 F_i= \frac{\pa W}{\pa \phi_i}
 \ee
and \be  D_i=g\sum_a \phi_i^{*}T^a \phi_i.\ee (We hope that the
reader does not confuse our notations $\Phi$ with $\phi_i$ and
$M_X$ (below) with mass of the  SM  Higgs  boson $H$).

Therefore, the general Higgs potential becomes \bea
V_{tot}&=&|\mu_\chi  +\alpha \Phi |^2|\chi^\prime|^2+ |\mu_\chi
+\alpha \Phi |^2|\chi|^2+|\alpha \chi
\chi^\prime -\mu^2|^2+\\
\nonumber & & + |\mu_\rho \rho|^2+|\mu_{\rho^\prime
}\rho^\prime|^2 + \frac{1}{2}\sum_{\alpha}|D_\alpha|^2+
V_{soft}.\eea
 It is easy to see that the fields
$\rho, \rho^\prime $ have settled down to their minimum, since the
first derivatives $\frac{\pa V_{tot}}{\pa \rho}, \frac{\pa
V_{tot}}{\pa \rho^\prime}$ are independent of $\chi, \chi^\prime,
\Phi$. This  means that the fields $\rho, \rho^\prime$ will stay
in their minimum independently of what the fields $\chi,
\chi^\prime, \Phi$ do. On the other hand, we are mainly interested
in what is happening above the electroweak scale, and hence we do
not take into account the dimensional Higgs  multiplets $\rho,
\rho^\prime$. Therefore, the Higgs scalar potential is  given by
\bea V_{inf}&=&|\mu_\chi +\alpha \Phi |^2|\chi^\prime|^2+
|\mu_\chi +\alpha \Phi |^2|\chi|^2+|\alpha \chi \chi^\prime
-\mu^2|^2+ \crn & & + \frac{1}{2} \left( g\sum_a\chi^* T^a
\chi\right)^2+\frac{1}{2} \left(g\sum_a\chi^{\prime*} T^a
\chi^\prime\right)^2. \label{Poten2} \eea
 Now, let us  make the change of variables
 \be
 \mu_\chi
+\alpha \Phi \equiv \beta S,\label{nha}\ee where $\bet$ is some
constant and $S$ is a new field.  Then, the Higgs potential
(\ref{Poten2}) can be rewritten as \bea V_{inf} &=& \beta^2
|S|^2\left( |\chi|^2+ |\chi^\prime|^2\right)+|\alpha \chi
\chi^\prime -\mu^2|^2 \crn && + \frac{1}{2} \left( g\sum_a\chi^*
T^a \chi\right)^2+\frac{1}{2} \left(g\sum_a\chi^{\prime*} T^a
\chi^\prime\right)^2. \eea

 When $D$ term vanishes along $D$ term direction, the potential
 contains only $F$ term ,it hence  can be written as
 \be
V_{inf}=\beta^2 |S|^2\left( |\chi|^2+
|\chi^\prime|^2\right)+|\alpha \chi
\chi^\prime-\mu^2|^2.\label{th271}
 \ee
From (\ref{th271}), it is clear that  $V_{inf}$ has an unique
supersymmetric minimum corresponding to
 \bea
 \langle  S  \rangle & = & 0, \crn
 M_X & \equiv &  \langle |\chi|\rangle  =  \langle
 |\chi^\prime|\rangle=\frac{\mu}{\sqrt{\alpha}}.
\label{th272} \eea  The ratio $\frac{\mu}{\sqrt{\alpha}}$ sets the
$U(1)_Z$ symmetry breaking $M_X$, but  Eq. (\ref{th272}) is global
minimum, and supersymmetry is not violated~\cite{GDvali}. Hence,
inflation can take  place but supersymmetry is not broken.

To finish this section, we emphasize  that our construction leads
to $F$ term inflation (see classification in ~\cite{Jen}).

 \section{Effective potential for inflation}
\label{effective}

 The above-proposed model belongs to the theories with spontaneous
 symmetry breaking. To make the predictions in good agreement with
 observed data, particularly
 with the most recently WMAP five-year analysis,  it is necessary to add the
 radiative corrections (see an example in
 ~\cite{chao}).  For
 studying radiative corrections
 in  theories with spontaneous symmetry breaking, the  method of effective
 potential~\cite{Coliman} is extremely useful. Let us apply
 the method for our inflationary scenario.

 The main idea of the chaotic inflation scenario was to study all
possible initial conditions in the Universe, including those which
describe the Universe outside of the state of thermal equilibrium,
and the scalar field outside of the minimum of potential
$V(\phi)$~\cite{lind383}.
 Now, let us assume chaotic initial conditions, and for this purpose,
 we denote the critical value for the inflaton field by
 $|S_c| \equiv \frac{\mu}{\sqrt{\alpha}}$.
  The initial
 value for the inflaton field is much greater than its critical
 value $S_c$.
 For $|S|> |S_c|$  the potential is
 very flat in the $|S|$ direction, and the $\chi, \chi^\prime $
 fields settle down to the local minimum of the potential,
 $\chi=\chi^\prime=0$, but it does not drive $S$ to its minimum
 value.  The universe is dominated by a nonvanishing vacuum energy density,
 $V_0^{\frac{1}{4}}=\mu$, which can lead to an exponential expanding,  inflation starts,
  and supersymmetry is
 broken. The inflaton field $S$ must have couplings to  matter
 field, which allow an universe to make the transition to Hot Big Bang
 cosmology at the end of inflation. These couplings will induce
 quantum corrections to potential $V$. Thus,
 we obtain some quantum correction to effective
 potential.  The good agreement between
the Coleman-Weinberg quartic potential and  the WMAP result has
been shown in ~\cite{so}.
   By the Coleman-Weinberg
 formula in \cite{Coliman}, at the one-loop level, the effective potential along the
 inflaton direction is given by
 \bea
 \Delta V=\frac{1}{64\pi^2}\sum_i ( -1)^F m_i^4
 \ln\left( \frac{m_i^2}{\Lambda^2}\right),
 \eea
where  $F= -1$  for the fermionic fields and
 $F=1$ for the bosonic fields. The coefficient $(-1)^F$ shows that bosons and fermions
 give opposite contributions.  The sum runs over each degree of
 freedom $i$ with mass $m_i$ and $\Lambda$ is a renormalization scale.

 Note that for $S > S_c$ there is no mass splitting inside the gauge
 supermultiplets. When $S$ falls below $S_c$,
 we can obtain nonvanishing contribution from the mass splitting of
 the $\widehat{\chi}, \widehat{\chi^\prime}$ superfields.
During the inflation only Higgs bosons and their fermionic
superpartners in the $\chi$ and $\chi^\prime $ triplets  give
contribution to the effective potential.  For the simplicity, we
assume the mass degeneracy among them.
 Therefore, the particle mass
 spectrum during $F$ term inflation contains: (i) six complex scalar
 fields with square mass  $\beta^2 |S|^2\pm \alpha \mu^2$. It means that
 they split into six pairs of real and pseudoscalar components.
 (ii) six fermionic fields   get mass $\beta |S|$.
  Hence,
the effective potential (along the inflationary trajectory $S
> S_c,\chi=\chi^\prime=0$) at the one loop has a form
 \bea
  V_{eff} (S)&=& \mu^4+\frac{3}{16\pi^2}\left[ 2 \beta^4 \frac{\mu^4}{\alpha^2}
  \ln\frac{\beta^2 |S|^2}{\Lambda^2}
 +\left( \beta^2 |S|^2+ \alpha \mu^2\right)^2
  \ln \left( 1+\frac{\alpha\mu^2}{\beta^2 |S|^2}\right)+ \right.
  \crn && \left.
   + \left( \beta ^2|S|^2- \alpha \mu^2\right)^2 \ln \left( 1-\frac{\alpha\mu^2}{\beta^2
   |S|^2}\right)\right]\label{Veff}
  \eea
 It is to be noted that  for  $S > S_c $, the universe is dominated by the false
 vacuum  energy $\mu^4$. When $S$ field drops to $S_c$, then the
 GUT phase transition happens. At the end of inflation, the inflaton field does not
 need to coincide  with the GUT phase transition. The end of inflation
 can be supposed to be on a region of the potential which
 satisfies the flatness conditions (see, for example,
 ~\cite{lythrioto})
 \bea
 \epsilon  \ll1 , \
 \eta   \ll 1,
 \label{cond1}\eea
 where  we have used the conventional notations
 \bea
 \epsilon & \equiv &\frac{M_P^2}{16 \pi}\left(\frac{V^\prime}{V}
 \right)^2,\   \eta \equiv \frac{M_P^2}{8\pi} \frac{V^{\prime \prime}}{V
 }
\label{cond11},\eea
 where  primes denote a derivative with respective  to $S$.

\section{ $F$-term inflation contribution to the CMB temperature anisotropy}
\label{finf}

 Having explored the model, we move on to comparison
with observational data. For this purpose, the parameters were
chosen to give a good representation of the observational data.
The exception to this is the microwave background on large angle
scales. The crucial COBE observations are the earliest to
interpret in the context of inflation, and they are also more or
less definitive because, on large angle scales, their accuracy is
limited not by instrument noise, but by the statistical
uncertainty known as cosmic variance, arising from our having only
a single microwave sky to look at. The standard approximation
technique for analyzing inflation is the slow-roll approximation.
This section is devoted to the observable parameters of the
models, namely the slow-roll ones: $\ep$ and $\eta$.
 The first condition in (\ref{cond1}): $\epsilon \ll1$ indicates that the density $\rho$
 is close to $V$ and is slowly varying. As a result,  the  Hubble parameter  $H$ is
 slowly varying, which implies that one can write $a\propto
 e^{Ht}$ at least over a Hubble time or so. The second condition $\eta \ll 1$
 is a consequence of the first condition plus the slow-roll
 approximation. The  conditional   phase may end before the GUT
 transition if the flatness conditions (\ref{cond1})  are violated at some
  point  $S > S_c$.

 For the sake of convenience, let us denote a dimensionless
variable \be y \equiv  \frac{\beta |S|}{\alpha S_c
}.\label{th721}\ee
 Then the expressions for $\ep$ and $\eta$
 in (\ref{cond11})
 can be written in the form
\bea
 \epsilon &=&\left(\frac{3M_P}{4\pi^2 M_X}\right)^2\frac{1}{16
 \pi}\left\{
  \frac{\beta}{\alpha y}\left( \frac{\beta^4}{\alpha^2}-\alpha^2\right)+\beta \alpha
  \left[ y\left( y^2-1\right)\ln \left(
  1-\frac{1}{y^2}\right)+\right.\right.
 \crn && \left. \left.
 + y\left( y^2+1\right)\ln \left(
 1+\frac{1}{y^2}\right)\right]\right\}^2,
 \crn
  \eta &=&\left(\frac{ M_P}{4\pi M_X}\right)^2\frac{3\beta}{2\pi \alpha}
  \left\{\frac{-\beta}{\alpha y^2}\left(\frac{\beta^4}{\alpha^2}-\alpha^2
  \right)+\right.
  \crn &&  \left.+ \alpha \beta \left [ (3y^2+1)\ln\left(1+\frac{1}{y^2} \right)+(3y^2-1)
  \ln\left(1-\frac{1}{y^2} \right) \right] \right\}. \label{cond2}\eea
  If we  impose the condition $\alpha =\beta$,  which means that $|\Phi| \approx |S| \gg \mu_\chi$,
   then the expressions in
   (\ref{cond2})   become
   \bea
 \epsilon &=&\left(\frac{3\alpha^2 M_P}{4\pi^2 M_X}\right)^2\frac{1}{16
 \pi}
  \left[
  y\left( y^2-1\right)\ln \left( 1-\frac{1}{y^2}\right)+
 \right.\crn && \left.
 + y\left( y^2+1\right)\ln \left( 1+\frac{1}{y^2}\right)\right]^2, \crn
  \eta &=&\left(\frac{\alpha M_P}{4\pi M_X}\right)^2\frac{3}{2\pi }
  \left[(3y^2+1)\ln\left(1+\frac{1}{y^2} \right)+(3y^2-1)\ln\left(1-\frac{1}{y^2} \right)
   \right] \label{cond3}\eea

 The  slow-roll parameter $\eta$ tends
  to infinity when $y$ approaches 1, so that inflation
 ends as $y$ turns to 1. We will find the value of the scalar coupling
 $\alpha$, the scale $M_{GUT},$ and the scale of $M_X$  which leads to the
 successful inflation. If we take the value of $M_X$ lying in
  $10^{15} - 10^{16}$
 GeV  and the value of $\alpha$ lying in $15  - 19$, the inflation
 does not happen.
   However, if we take the value of $\alpha$ being in
 $10^{-2} - 10^{-3}$ with the value of $M_X$ belonging to
 $10^{15} - 10^{16}$  GeV,
 then both $\epsilon$  and $\eta$ do not reach to unity.
 When $|S|$ falls below $S_c$, the slow-roll conditions are
 violated, and inflation stops. The unwanted monopoles have been
 inflated away.

 Since along the valley the dynamics satisfies the slow-roll
 conditions, we also easily evaluate the cosmic background
 radiation quadrupole anisotropy. The temperature fluctuations
 in the cosmic microwave background
 radiation (CMBR) are proportional to the density perturbations which
 were produced in the very early universe and
 lead to structure formation
 \be
 \frac{\delta T}{T}= \frac{1}{3}\frac{\delta \rho}{\rho}.
 \ee
In our case, the $F$-term hybrid inflationary scenario, cosmic
strings form at
  the end of inflation. Therefore, we must calculate the contribution to $\frac{\delta
 \rho}{\rho}$ and $\frac{\delta T}{T}$ from both inflation and
 cosmic strings. It means that both  inflation and cosmic string are
 the part of one scenario, temperature fluctuations in the CMBR
 are a result of the quadratic sum of the temperature fluctuations
 from inflationary perturbation and cosmic strings. The
perturbations from inflation and cosmic string are uncorrelated
and
 add up independently~\cite{Jen}: \be
\left(\frac{\delta T}{T}\right)_{tot} = \sqrt{\left( \frac{\delta
T}{T}\right)^2_{inf}+\left( \frac{\delta T}{T}\right)^2_{cs}}. \ee
The inflation contribution to the the CMB temperature anisotropy
is \be \left( \frac{\delta
T}{T}\right)_{inf}=\frac{1}{12\sqrt{5}\pi M_P^3}
\frac{V^{\frac{3}{2}}}{V^\prime}|_{S=S_Q},\label{for1}\ee where
the subscript Q denotes the time observable scale leaves the
horizon. It means that the right-hand side of (\ref{for1}) must be
evaluated at the epoch of horizon exit. The spectral index of
density perturbations can also be expressed in terms of the slow
parameters: \be n=1-6\epsilon +2\eta. \label{spectral1}\ee

It can be evaluated at any scale.  The number of e-foldings
between two values of the inflaton field $S_{end}$ and $S_Q$ is
given by \be N(S_Q)\simeq
\frac{8\pi}{M_P^2}\int^{S_{Q}}_{S_{end}}\frac{V}{V^\prime} dS,
\label{for2}\ee where $S_{Q}$ is called `\emph{`the value of $S$
at the cosmological scale to leave the horizon}''.  Taking
cosmologically interesting scales to go from $H_0^{-1}\sim 10^4$
Mpc to 1 Mpc gives a range \cite{book} $\Delta N_Q =\ln 10^4 \sim
1 \ \textrm{Mpc}$ and the value of $ N(S_Q)$ lies in the range 50
to 25, corresponding to an interrupted Hot Big Bang, and to a
delayed Hot Big Bang preceded by one bout of thermal inflation.

 If $S$ is sufficiently greater than $S_c$ and
$\alpha =\beta$, then the effective potential given in
(\ref{Veff}) reduces to a simple form \be V_{eff}(S)= \mu^4 \left[
1+\frac{3\alpha^2}{16\pi^2}\left( 2\ln\frac{\alpha^2
S^2}{\Lambda^2}+\frac{3}{2}\right)\right]\label{for3} \ee This
potential occurs as part of hybrid inflation model. If we assume
that the loop correction dominates the slope, and use
(\ref{cond11}), the slow-roll parameters are \bea \epsilon & = &
\frac{9 M_P^2}{256 \pi^5}\frac{\alpha ^4}{S^2}, \crn \eta & = &
-\frac{3 M_P^2}{32 \pi^3}\frac{\alpha^2}{S^2}.\eea Here we assume
that inflation ends when $|\eta|=1$. The value of the inflation
field at the end of inflation is given by \be S_{end}^2=\frac{3
M_P^2 \alpha^2}{32 \pi^3}. \label{for4}\ee

 From Eqs. (\ref{for2}), (\ref{for3}), and (\ref{for4}),
 the value of $S$  at the cosmological scale to leave the horizon
  ($S_Q$), is given by
   \be
S_Q^2=\left[ 2N(S_Q)+1\right]\frac{3 \alpha^2 M_P^2}{32 \pi^3},
 \label{them1}\ee
 and the spectral index evaluated at this scale is as follows:
 \be
n_{N(S_Q)}=1-\left( \frac{9 \alpha^2}{4
\pi^2}+2\right)\frac{1}{2N(S_Q)+1}.\label{them1m}
 \ee

\begin{figure}[htbp]
\bc
\includegraphics[width=12cm,height=12cm]{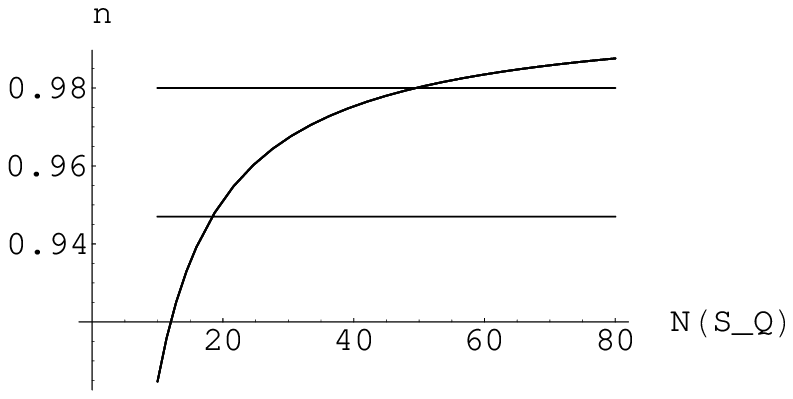}
\caption{}
\label{resubh1} \ec
\end{figure}

 The spectral index is a function of the coupling $\alpha$ and
 the e-folding number $N(S_Q)$. If we take the value of the
 coupling $\alpha $  smaller than $10^{-1}$, then the contribution
 of $\frac{\al^2}{4\pi^2}$ can be neglected. It means that
  the value of the spectral index $n$ is
 unchanged too much when the values of $\al$ varying from $10^{-1}$ to
 0.  As a specific example, fig.\ref{resubh1} represents the
 spectral index as a function of $N(S_Q)$ with  $\al  \in [10^{-1},
 10^{-7}]$. Precisely speaking: by  (\ref{them1m}), if $\alpha < 0.1$, the spectral index
 is not sensitive to it.

 Comparing with the WMAP collaboration five-year data~\cite{wmapdm}
 \be n = 0.960^{+0.014}_{-0.013},\label{th721m}\ee
 we obtain that the value of the e-folding number at the  cosmological scale to
 leave the horizon  ($S_Q$) lies in the range 25
 to 50 and $\al \in [10^{-1},10^{-7}]$.

\begin{figure}[htbp]
\bc
\includegraphics[width=12cm,height=12cm]{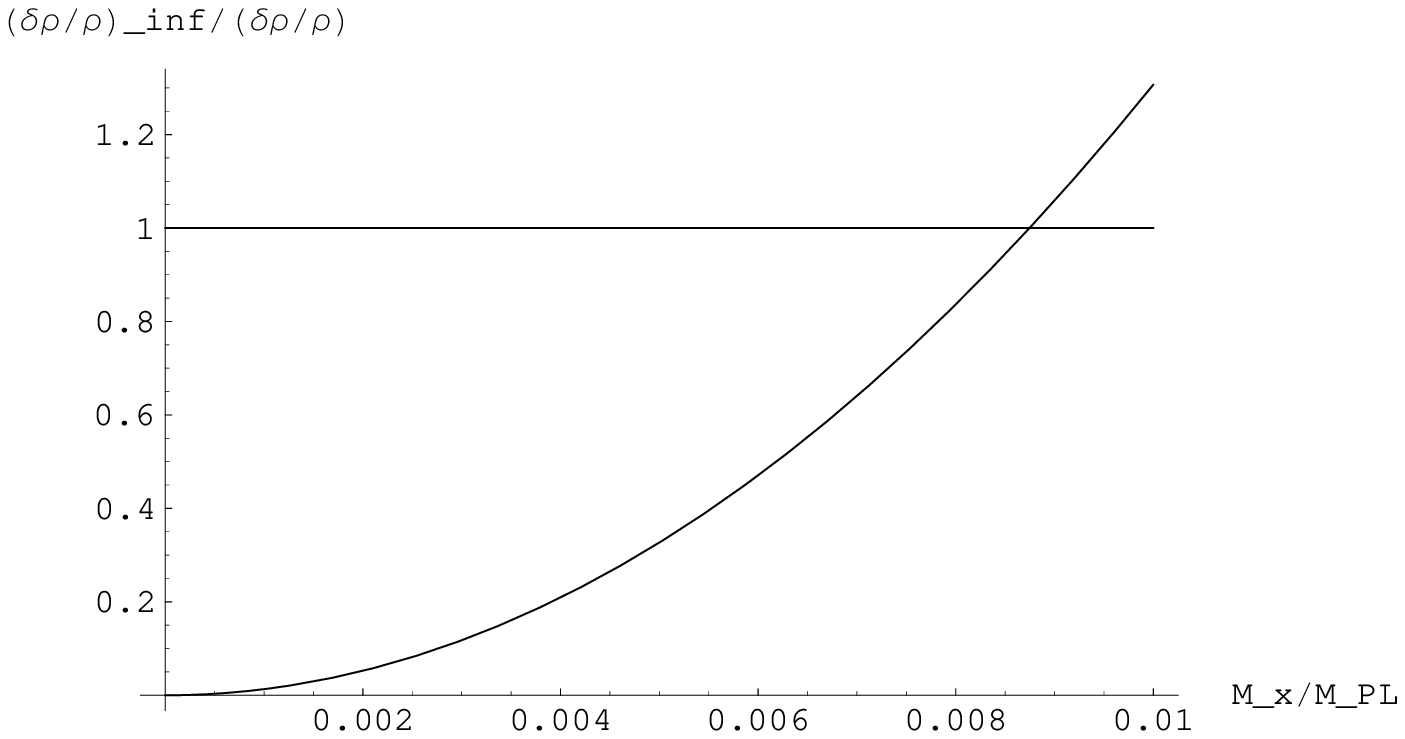}
\caption{}
\label{resubh2} \ec
\end{figure}

  From Eqs. (\ref{for1}) and (\ref{for3}), the inflation contribution
 to the  CMB temperature anisotropy is given by
 \be
\left( \frac{\delta T}{T} \right)_{inf}=
\frac{1}{36}\sqrt{\frac{3[2N(S)+1]}{10 \pi}}\left(
\frac{M_X}{M_P}\right)^2.
 \ee
 Taking $N(S)=50$, we can  see the temperature anisotropy as a
 function of $\frac{M_X}{M_P}$. In  Fig. \ref{resubh2}, we have
 plotted the inflationary contribution to quadrupole and have compared with the
 quadrupole measured by COBE ~\cite{cobeW}
 \be
\left( \frac{\delta T}{T} \right)_{COBE}= 6.6 \times 10^{-6}.
 \label{cobe}\ee
 We obtain that the inflationary contribution to quadrupole is 10\%  to
 100\%  if  ratio $\frac{M_X}{M_P}$ lies in the range from $ 10^{-3}$
 to $9 \times 10^{-3}$. Taking $M_P=1.22 \times 10^{19}$ GeV, we
 get a constraint for  $M_X \in[1.22 \times 10^{16}, 0.98 \times 10^{17}]
 $ GeV.

According to the data given by ~\cite{less10}, the string
contribution to quadrupole is less than 10\%. It means that
inflationary contribution to the mixed scenario with both
inflation and cosmic string is 90\%. In this case, we get the
ratio of $\frac{M_X}{M_P}=8 \times 10^{-3}$. Combining this data
and $M_P=1.22 \times 10^{19}$ GeV, it follows that the value of
$M_X=9.76 \times 10^{16} \textrm{GeV} \simeq 10^{17}$ GeV.

 Let us estimate the $\mu $ parameter. Substituting   $\al =10^{-2}$
  into  Eq.(\ref{th272})
 we get \be \mu =
\sqrt{\al}\ M_X\approx \sqrt{10^{-2}}\ M_X  \simeq 10^{16} \
\textrm{GeV}. \ee Thus, the parameter $\mu$ is smaller than the
mass of inflaton and
is much larger than $\mu_\chi$, which is in the TeV scale
(see~\cite{susyeco,higph,jhep2}). Hence our approximation $\al
\approx \bet$ is completely good.

It is worth reminding that  from  (\ref{them1}) the value of the
inflation field at the cosmological scale to leave the horizon
$S_Q = 1.12 \times 10^{17}$ GeV corresponds to the e-folding
number $N(S_Q)=50$ and $\al =10^{-2}$.

 It is interesting to note that chaotic
inflation driven by the $\phi^2$ potential is in good agreement
with the most recent five-year WMAP
 analysis. For the $\phi^4$ potential, the predictions
for spectral index $n$ and $r = 16\ \epsilon$  lie outside the
WMAP 95\% confidence level~\cite{chao}.

Note that  alternative $D$-term inflation, which exists only for
simple gauge group, is not available  in the framework of 3-3-1
models.

To finish this section, we remind  the reader that to satisfy the
five-year WMAP data given in $(\ref{th721m})$, the number of
e-foldings $N(S_Q)$ must be dropped much below 50. Hence, one
cannot resolve the horizon/flatness problems of Big Bang
cosmology. In addition a value of the coupling $\al$ is not
sensitive to the value of spectral index $n$. On the other hand,
in supersymmetric theories based on supergravity, there is a well
known problem  that  $\eta =1$ due to the supergravity
corrections, thereby violating one of the slow-roll conditions  $
\eta \ll 1$. This is the  so-called $\eta$ problem. To make good
one's shortcomings, we will consider $F$-term hybrid standard
inflation with minimal K$\ddot{a}$hler potential.

\section{Standard $F$-term inflation with minimal K\"{a}hler potential}
\label{sec6}
 The standard $F$-term inflation with  K\"{a}hler potential is
defined by the super-potential
 \be
W_{stand}(\Phi,\chi,\chi^\prime)= \al \widehat{S}
\left(\widehat{\chi} \widehat{\chi^\prime} -M_X^2
\right),\label{sugra2}\ee and  the  K\"{a}hler potential, in
general case, is determined from the Lagrangian \cite{susy} \be
\mathcal{L} = \int d^2 \theta d^2 \bar{\theta} K (\Phi^i, \Phi^{*
i}). \ee The supergravity potential  including  $F$-term has the
form \cite{crem1}
 \bea V & = & e^{K/m_P^2}\left[ (K^{-1})_i^j F^iF_j - 3\fr{| W|^2}{m_P^2}\right]
 + \fr{g^2}{2} \textrm{Re} f^{-1}_{AB} D^A D^B,
\label{sugra11} \eea where \be F^i = W^i + K^i\fr{ W}{m_P^2}\ee
and \be D^i =  K^i(T^A)_i^j \phi_j + \xi^A.\ee
 The scalar potential is given by~\cite{Jean3} \be V_F=
e^{K/m_P^2}\left[ \sum_\al \left| \fr{\pa W}{\pa \phi_\al} +
\fr{\phi^* W}{m_P^2}\right|^2 - 3\fr{|
W|^2}{m_P^2}\right]\label{sugra1} \ee where the sum is over all
fields.

 Substituting
 Eq.(\ref{sugra2}) into Eq.(\ref{sugra1}) with  the minimal K\"{a}hler
 potential, keeping in mind  that
  $K = \sum_\al \left|\phi_\al \right|^2$, we obtain the scalar potential
 given in the following form
 \bea V^m_F & = & 2 \al^2 S^2 \phi^2 \left [1 + \fr{S^2 +
 2\phi^2}{m_P^2} + \fr{(S^2 +
 2\phi^2)^2}{2 m_P^4} \right]+\crn
&&+\al^2(\phi^2 -M_X^2)^2\left( 1 + 2\fr{\phi^2}{m_P^2} +
\fr{S^4}{2 m_P^4} + 2\fr{\phi^4}{m_P^4}\right) + \cdot \cdot
\cdot\label{sugra3} \eea Here we have assumed that $|\phi|^2 =
|\phi^\prime|^2$ . Note that the first line in the right-hand side
of  Eq. (\ref{sugra3}) contains the following terms: \be
\fr{S^2}{m_P^2} + \fr{S^4}{2 m_P^4}, \label{sugra4}\ee which were
neglected in ~\cite{bastero}; they  will give a correction to
mass of the Higgs $\phi$.
 Let us consider how  does this factor change the result. As we know, the
slow-roll parameter  is defined as \bea \eta = m_p^2 \left(
\frac{V^{\prime \prime}}{V}\right)  \eea

The prime refers to derivative with respect to $S$. With $V$ given
in (\ref{sugra3}), the supergravity scalar potential for $S>S_c$
is given by \be V_o=\al^2 M_X^4 + \frac{\al^2 M_X^4}{2m_p^4}S^4.
\label{sugra4} \ee There is no  mass term for the inflaton field
$S$ in (\ref{sugra4}). Hence, we have to calculate the second
derivative of $V$:  $ V^{\prime \prime }\simeq \frac{\al^2
M_X^4}{2 m_p^4} S^2$. This yields  $\eta = \frac{1}{2 m_p^2} S^2
\ll 1$. It means that the $\eta$-problem is solved.

 We would like to say again that for $S>S_c$, the energy density is
dominated by the vacuum energy $\al^2 M_X^4$, which therefore
leads to an exponentially expanding (inflationary) universe. The
potential given in (\ref{sugra4}) does not contain a term which
can drive $S$ to its minimum value. It means that we have to
consider the effective potential. Taking into account  one-loop
correction, we can write the potential in the form
 \be V= V_0+ V_{oneloop}
\ee with the radiative correction given by \be V_{oneloop}=
V_{nonsugra}+V_{sugra}, \label{Potental}\ee
 where $V_{nonsugra}$ is the effective potential which is  obtained
  from one-loop correction
 without  the  K\"{a}hler potential:
\be V_{nonsugra}=\frac{3 \al^4 M_X^4}{32 \pi^2}\left[ (x^2-1)^2
\ln \left(1-\frac{1}{x^2} \right)+(x^2+1)^2 \ln
\left(1+\frac{1}{x^2} \right)+2\ln\frac{\al^2 M_X^2
S^2}{\Lambda^2}\right], \label{su1}\ee and $V_{sugra}$ is the
supergravity correction: \bea V_{sugra}&=& \frac{3\al^4 M_X^4}{32
\pi^2}\left[(x^2-1)^2 \ln [1+\zeta (x^2-1)] + (x^2+1)^2 \ln
[1+\zeta (x^2+1)]  \right]+ \crn  && + 2 \zeta (x^2-1)^3 \ln
\left(1-\frac{1}{x^2} \right)+2 \zeta (x^2+1)^3 \ln
\left(1+\frac{1}{x^2} \right),\label{su2} \eea where
$x=\frac{|S|}{M}$ and $\zeta= \frac{M_X^2}{M_P^2}$. We emphasize
that in obtaining  the  potential given in (\ref{su1}) and
(\ref{su2}),  the quartic terms   of $\zeta ^2$ were neglected.

  The slow-roll parameters are given by
 \bea
 \epsilon & \equiv &\frac{M_P^2}{16 \pi}\left(\frac{V_0^\prime+
 V_{nonsugra}^\prime+V_{sugra}^\prime}{V}
 \right)^2 =\left(\frac{3 \al^2  M_P}{8 \pi^2 M_X}\right)^2\frac{1}{16 \pi}
 \left(\varepsilon_o+\varepsilon_{nonsugra}+\varepsilon_{sugra} \right)^2,
  \crn\   \eta &\equiv& \frac{M_P^2}{8\pi} \frac{V_o^{\prime \prime}+
  V_{nonsugra}^{\prime \prime}+V_{sugra}^{\prime \prime}}{V
 }= \eta_0 +\eta_{nonsugra}+\eta_{sugra},
\label{cond11t}  \eea where \bea \varepsilon_0 &=& \left( \frac{8
\pi^2}{3\al^2} \right)^2 2 \zeta^2 x^3, \crn
\varepsilon_{nonsugra}&=& x\left[ (x^2-1) \ln
\left(1-\frac{1}{x^2} \right)+(x^2+1) \ln \left(1+\frac{1}{x^2}
\right)\right], \crn \varepsilon_{sugra}&=& x\left[ (x^2-1) \ln
\left(1+\zeta(x^2-1) \right)+(x^2+1) \ln \left(1+\zeta(x^2+1)
\right)+\zeta \left( x^4+1\right)+\right.\crn &&+ \left.3 \zeta
(x^2-1)^2\ln \left(1-\frac{1}{x^2} \right) +3 \zeta (x^2+1)^2\ln
\left(1+\frac{1}{x^2} \right)-4\zeta \right]\eea and the $\eta$
parameters are given by \bea \eta_0 &=& \frac{3}{4 \pi}\zeta^2
x^2, \crn \eta_{nonsugra}&=&\left(\frac{ \al M_p}{4 \pi M_X}
\right)^2 \frac{3}{4\pi}\left[ (3x^2-1) \ln \left(1- \frac{1}{x^2}
\right)+ (3x^2+1) \ln \left(1+ \frac{1}{x^2} \right)\right], \crn
\eta_{sugra}&=& \left(\frac{\al M_p}{4 \pi M_X} \right)^2
\frac{3}{4\pi}\left[ (3x^2-1) \ln \left(1+\zeta \left(x^2-1
\right)\right)+ (3x^2+1) \ln \left(1+ \zeta \left(x^2+1 \right)
\right)+ \right.\crn && + \left. \zeta \left( 9 x^4 +1 +3(x^2-1)^2
\ln \left( 1-\frac{1}{x^2}\right) + 3(x^2+1)^2 \ln \left(
1+\frac{1}{x^2}\right)+\right. \right.\crn &&+ \left. \left.12x^2
\left[ (x^2-1)\ln \left( 1-\frac{1}{x^2}\right)+ (x^2+1)\ln \left(
1+\frac{1}{x^2}\right) \right] -16\right)\right].\eea

\begin{figure}[htbp]
\bc
\includegraphics[width=12cm,height=12cm]{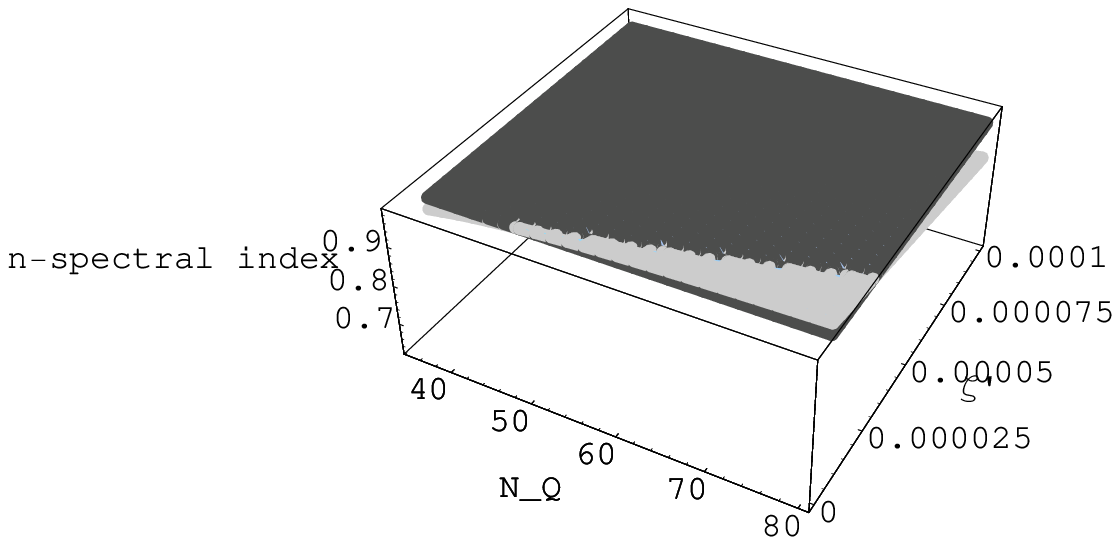}
\caption{}
\label{specsugra1}
\ec
\end{figure}

\begin{figure}[htbp]
\includegraphics[width=12cm,height=12cm]{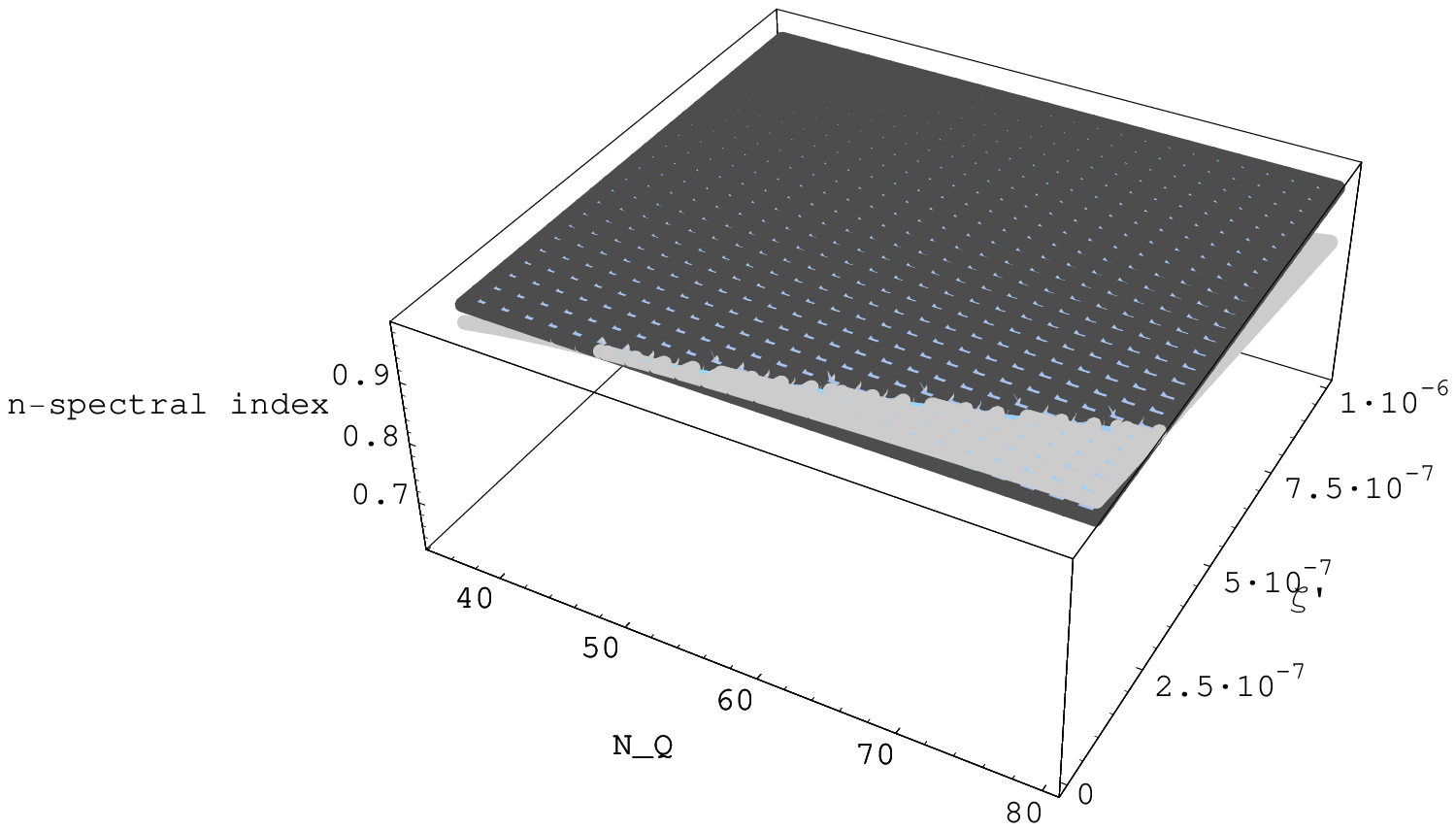}
\caption{}
\label{specsugra2}.
\end{figure}

\begin{figure}[htbp]
\includegraphics[width=12cm,height=12cm]{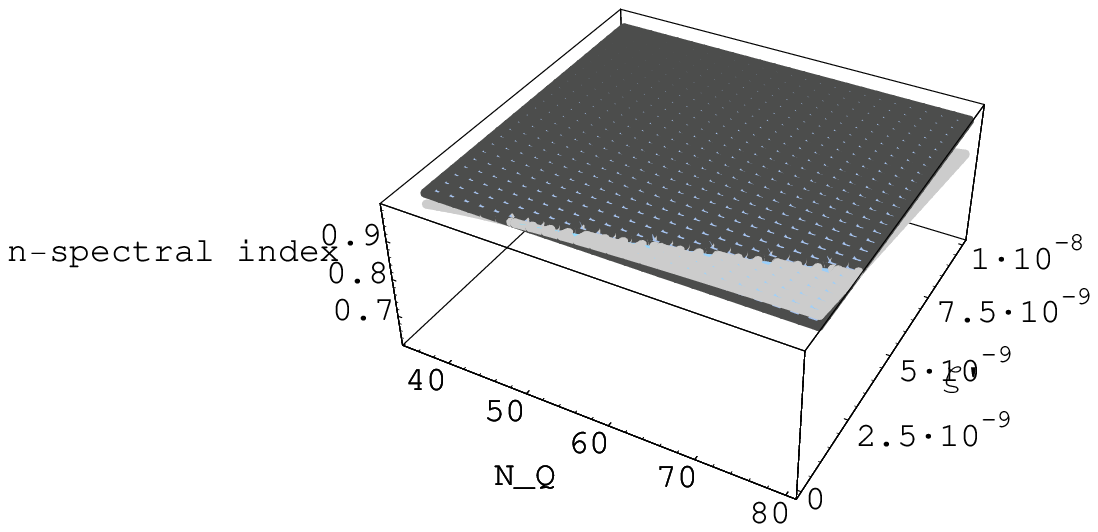}
\caption{}
\label{specsugra3}.
\end{figure}

\begin{figure}[htbp]
\includegraphics[width=12cm,height=12cm]{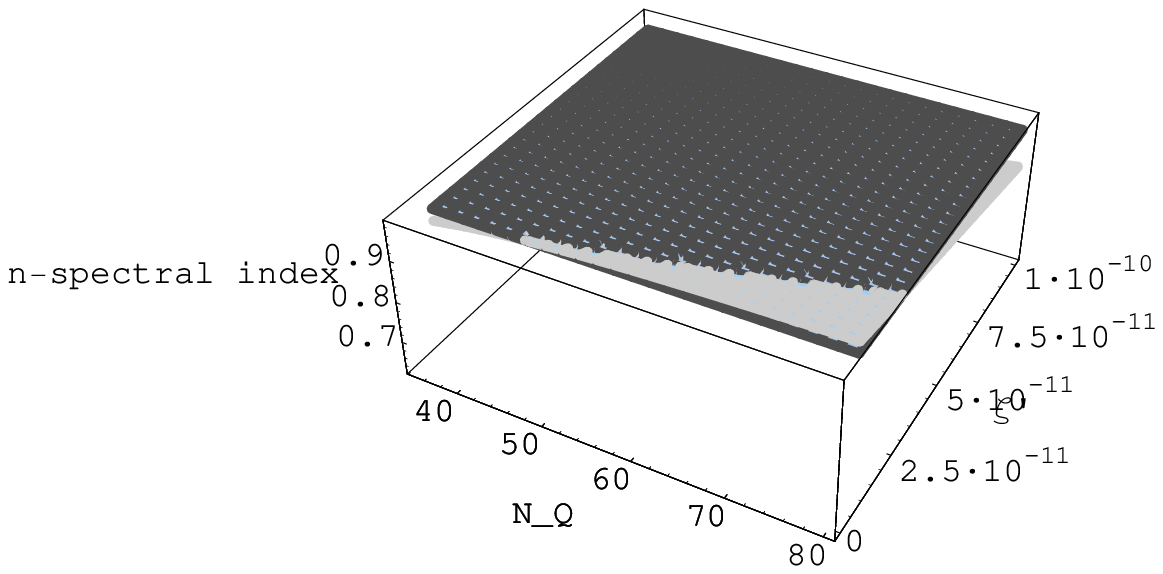}
\caption{}
\label{specsugrah}.
\end{figure}

In the  approximation of the first order of $\zeta$, the slow-roll
parameters  reduce to  a simple form such as
 \bea
 \epsilon &\simeq&\left(\frac{3 \al^2 M_P }{8 \pi^2 M_X}\right)^2\frac{1}{16 \pi}
 \left\{
 \frac{1}{x}+ \zeta x\left[ 3(x^4+1)\left(
 1-\frac{1}{x^4}\right)+8\right]
 \right\}^2\simeq \crn & \simeq & \left(\frac{3 \al^2 M_P }{8 \pi^2 M_X}\right)^2\frac{1}{16 \pi}
 \left\{
 \left(\frac{1}{x}\right)^2+ 2\zeta \left[ 3(x^4+1)\left(
 1-\frac{1}{x^4}\right)+8\right]
 \right\}, \crn \eta & \simeq& \left(\frac{\al M_P}{4 \pi M_X} \right)^2
\frac{3}{4\pi}\left\{-\frac{1}{x^2} +\zeta \left[3\left(5x^4-1
\right) +8\right]\right\}.
 \eea
 The spectral index $n$ given in (\ref{spectral1}) can be written
 as
 \be
n=1-\frac{3\al^2}{512 \pi^3 \zeta x^4}
\left[x^2(16+9\al^2)-54\al^2 \zeta +6x^8(-40+9 \al^2)\zeta
+16x^4(-5+9\al^2)\zeta \right] \label{spesugra} \ee In order to
estimate the value of $n$, we have to evaluate the field value $x$
at the $N(S_Q)$ e-folding number. Assuming that at the end of
inflation, $x_{end}=1$, and using the definition of $N(S_Q)$ given
in (\ref{for2}), one  can get an approximation  \be x\simeq
\frac{3\al^2 N(S_Q)}{16 \pi^2\zeta}. \label{for5} \ee Substituting
(\ref{for5}) into (\ref{spesugra}), we obtain the value $n$ as a
function of $N(S_Q), \zeta$, and $ \al$. The predicted value of
spectral index is plotted in Figs. \ref{specsugra1},
\ref{specsugra2}, \ref{specsugra3}, and \ref{specsugrah}.

Comparing the results from  Figs. \ref{specsugra1},
\ref{specsugra2}, \ref{specsugra3}, and \ref{specsugrah} with the
WMAP data given in (\ref{th721m}), the following conclusions are
in order: \ben
\item The value of e-folding number $N_Q$ must be larger than 45.
\item The  constraints for the value of  coupling $\al$ and  the parameter $\zeta$
are followed and presented in Table \ref{thf}: \een

\begin{table}[h]
\caption{
  Bounds on the parameter $\zeta$ and coupling $\al$ followed by the  WMAP data .}
\bc
\begin{tabular}{|c|c|c|c|c|}
  \hline
  $\al$ & $10^{-3}$ & $ 10^{-4}$  &  $10^{-5}$  &  $10^{-6}$ \\
  \hline
  $\zeta$ &  $25\times 10^{-6}$ &  $25 \times 10^{-7}$ &  $ 25 \times 10^{-9}$
  &   $3 \times 10^{-11}$\\
  \hline
\end{tabular}\label{thf}
\ec
\end{table}

 On the other hand,  requiring that the non adiabatic string
contribution to the quadrupole to be less than 10\%, we come to
conclusion that the value $M_X$ must satisfy \cite{Jean3} \be M_X
< 2.3 \times 10^{15} \sqrt{\frac{9/y}{\theta(\beta)}} \simeq 1.2
\times 10^{15}\ \textrm{GeV} \label{cosmic1}\ee or
 \be
\zeta=\frac{M_X^2}{M_P^2} < 10^{-8}.
 \label{thf2}\ee

\begin{figure}[htbp]
\bc
\includegraphics[width=12cm,height=12cm]{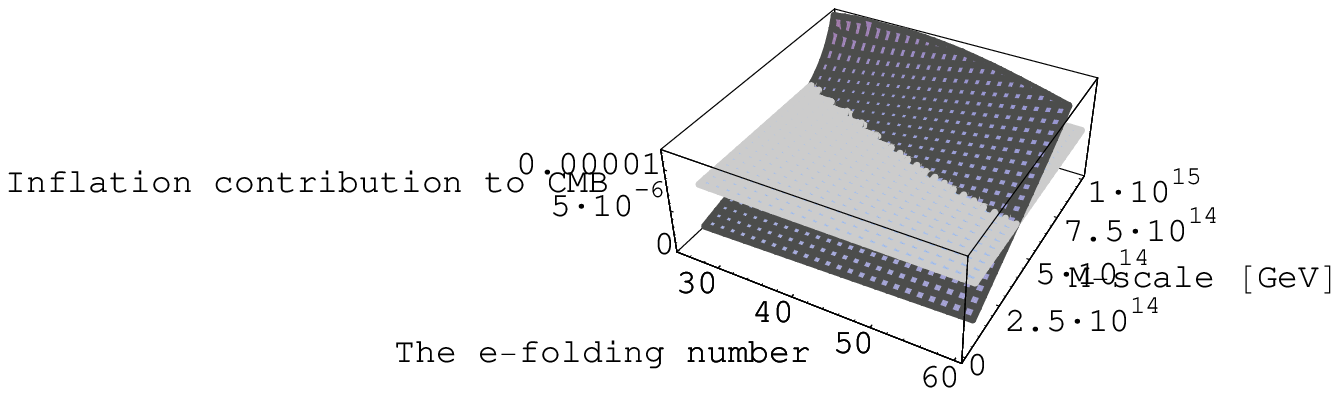}
\caption{}
\label{specsugra4} \ec
\end{figure}

 Combining Eq. (\ref{thf2}) with the constraint given in
table \ref{thf}, it follows  that the Higgs coupling $\al$ must be
smaller than $10^{-4}$. Let us  consider the inflation
contribution \textbf{to} the CMB temperature anisotropy  in a
specific case $\al = 10^{-6}$. Using the formulae given in
(\ref{for1}) with the potential given in (\ref{Potental}), we
obtain the inflation contribution to the CMB temperature as a
function of the number of e-folding $N(S_Q)$ and $M_X$,  which is
illustrated in Fig. \ref{specsugra4}.

From Fig. (\ref{specsugra4}),  we conclude that if the inflation
contribution to quadrupole is 90\%, the scale $M_X$ must be
equalled $3.5\times 10^{14}\  \textrm{GeV} $.

\section{Summary and conclusions}
\label{conclusion}

In the present paper we have constructed the supersymmetric
economical 3-3-1  model which can be consistent with cosmology.
Thanks to existence of two VEVs ($w, w^\prime$) which are singlets
under the SM gauge group, the derived model contains the hybrid
$F$-term inflation. Hence,  we have made a success of constructing
the model with inflation.  The effective potential was derived and
it has a global minimum. By the standard procedure, the slow-roll
parameters were calculated.

We have displayed the possible range of values for the
inflationary parameter in the model under consideration: \ben
\item
 From the analysis  of the inflation contribution to the
temperature fluctuations measured
 by the COBE, we see that the ratio $M_X/M_P$ depends on $N(S_Q)$ and
 the inflation contribution
to quadrupole. Taking $N(S_Q) =50$ and comparing with the COBE
data, we obtain the following results: (i) a range of $M_X$: $M_X
\in [1.22 \times 10^{16}\  \textrm{GeV} -  0.98 \times 10^{17}$
GeV], (ii) the parameter $\mu \simeq 10^{16} \ \textrm{GeV} $,
(iii)  and during inflation $S \simeq 10^{17}$ GeV.
\item Comparing with the five-year WMAP  data, we have shown that the
number of e-folding $N_Q$ is in the range of $25 - 50$. It means
that the cosmologically interesting scale, in which $H_0^{-1}\in
[10^4  - 1$ ] Mpc, requires a limit \cite{book}: $\Delta N_Q =\ln
10^4 \sim 1 \ \textrm{Mpc}$. This result is in good agreement with
estimation from previous works by other authors.
\item  In the usually accepted parameter space, the coupling
$\alpha$ varies in the range: $10^{-7} - 2 \times 10^{-1} $. The
coupling constant is not so small, and this is common character of
the supersymmetric unified models, to which the model under
consideration belongs.

\item The spectral index $n$ is approximately 0.98 if the e-folding
number is taken:  $N(S_Q)=50$. This value is a little bit
unsuitable to  the five-year WMAP data. \een

The scenario for large-scale structure formation implied by the
model is a mixed scenario for inflation and cosmic string.

 In the case of the standard hybrid inflation, the cosmic
string contribution is proportional to the square of $M_X$. So if
 $ M_X$ increases, both cosmic string and inflation contributions increase.
Hence, it is difficult to get  the cosmic string contribution less
than 10\% and inflation contribution 90\% at the same scale $M_X$,
namely, to get cosmic string contribution to quadrupole
 less than 10\%, the values $M_X$ must be satisfied the condition given
in (\ref{cosmic1}), i.e. $M_X \simeq 10^{15}$ GeV. However, in our
case, to get the inflation contribution to quadrupole is 90\%, the
value $M_X \simeq 10^{17}$ GeV. On the other hand, to get spectral
index $n$ smaller than $0.98$, the e-folding number must lie in
the range $[25 - 50]$. These values of e-folding number and the
spectral index $n$ is not only suitable to the WMAP data but also
cannot solve the horizon problem.

In the case of standard model with the  minimal K\"{a}hler
potential, our results are: \ben \item Assuming that the parameter
$\zeta = \frac{M_X^2}{M_P^2}$ is very small, we obtain the general
expression of the spectral indexes which is separated into two
parts. One part is gained from the standard hybrid inflation
potential and the other part is derived from the loop correction
of the  minimal K\"{a}hler potential. \item The inflation
contribution to quadrupole depends completely  on $M_X$. At the
value of $M_X =3.5 \times 10^{14}$ GeV, we obtain that the
inflation contribution to quadrupole is 90\%. This value is
suitable for the cosmic string contribution, which, according to
(\ref{cosmic1}), is less than 10\%.
\item The spectral index derived from  the  WMAP5 data ($n=0.96$),
can be justified if the e-folding number belongs to the range $[45
- 65]$.  With this  e-folding number, the horizon problem can be
solved.
\item From constraint of cosmic string contribution to
quardrupole, we obtain that  the Higgs coupling $\al$ must be
smaller than $10^{-4}$ GeV.\een

It is worth noting that the inflationary scenario is not available
for the non-supersymmetric economical 3-3-1 model due to the lack
of the  Higgs boson with necessary property. In general, analysis
in the present paper is available for other supersymmetric 3-3-1
models. However,  due to their large Higgs content, the analysis
will only be approximate.

In the present paper,  a new interesting property of the
supersymmetric economical  3-3-1 model that it can be extended to
describe the early universe  was found. The above mentioned model
has very nice advantage that its Higgs sector is minimal. Hence
its eigenmasses and eigenstates can be found exactly.

One of the criteria for the inflationary scenario, beside
providing the predictions in good  agreement with observations of
the microwave background and large-scale structure formation,  is
an explanation of the origin of the observed baryon asymmetry. For
this aim,  we note that the economical 3-3-1 model contains {\it
the lepton number violating interactions} at the tree level
through the SM gauge bosons such as the neutral $Z$ and the
charged $W^\pm$ bosons ~\cite{haihiggs}. The above  property is
the unique character of the economical version.

 The authors would like to thank T. Inami for help and useful
 lectures on inflation during his visit to Institute of Physics, Hanoi.
 One of the authors (H. N. L.)  would like to thank the CERN Theory
 Division and the  LAPTH, Annecy, France
 for financial support of his visit where this work was  completed.
 This work was supported in part
by the National Foundation for Science and Technology Development
(NAFOSTED)  under grant  No: 103.01.16.09.\\[0.3cm]

 \vs

  \vs

 \vs

 \vs

 \vs

 \vs

 \vs

\newpage

\textbf{Figure captions }

 Fig 1:  The spectral index $n $ as a
function of the number of e-folding $N(S_Q)$. Two lines present
the bound of the spectral index by WMAP collaboration five year
data given in (\ref{th721m})

Fig 2: The inflationary contribution to the quadrupole as a
function of $\frac{M_X}{M_P}$

Fig 3: The spectral index $n $ as a function of the number of
e-folding $N(S_Q)$ and $\zeta$, taking $\al=10^{-3}$. The gray
plane presents the bound of the spectral index by the five-year
WMAP data given in (\ref{th721m})

Fig 4: The spectral index $n $ as a function of the number of
e-folding $N(S_Q)$ and $\zeta$, taking $\al=10^{-4}$. The gray
plane presents the bound of the spectral index by the  five-year
WMAP  data given in (\ref{th721m})

Fig 5: The spectral index $n $ as a function of the number of
e-folding $N(S_Q)$ and $\zeta$, taking $\al=10^{-5}$.  The gray
plane presents the bound of the spectral index by the  five-year
WMAP  data given in (\ref{th721m})

Fig 6: The spectral index $n $ as a function of the number of
e-folding $N(S_Q)$ and $\zeta$, taking $\al=10^{-6}$. The gray
plane presents the bound of the spectral index by the  five-year
WMAP data given in (\ref{th721m})

Fig 7: The inflation contribution to CMB temperature  as a
function of the number of e-folding $N(S_Q)$ and $M_X$ in the case
$\al=10^{-6}$ . The light plane presents 90\% of the quadrupole
measured by COBE given in (\ref{cobe})
\end{document}